\documentclass[aps,preprint,singlecolumn,superscriptaddress,showpacs]{revtex4-1}
\usepackage{graphicx,color,xfrac}
\usepackage{bm}
\usepackage{amsmath}

\def \na213{Na$_2$IrO$_3$}
\def \li213{Li$_2$IrO$_3$}

\def \A213{$A_2$IrO$_3$}

\def \nhh{$^\mathcal{H}\langle N \rangle$-\li213}
\def \1hh{$^\mathcal{H}\langle 1 \rangle$-\li213}
\def \0hh{$^\mathcal{H}\langle 0 \rangle$-\li213}

\def \ioi{Ir-O$_2$-Ir}

\begin{document}

\title{Realization of a three-dimensional spin-anisotropic harmonic honeycomb iridate}

\author{K. A. Modic}
\affiliation{Los Alamos National Laboratory, Los Alamos, NM 87545, USA}
\affiliation{Department of Physics, University of Texas, Austin, TX 78712, USA}

\author{Tess E. Smidt}
\affiliation{Materials Science Division, Lawrence Berkeley National Laboratory, Berkeley, California 94720, USA}
\affiliation{Department of Physics, University of California, Berkeley, California 94720, USA}

\author{Itamar Kimchi}
\affiliation{Department of Physics, University of California, Berkeley, California 94720, USA}

\author{Nicholas P. Breznay}
\affiliation{Materials Science Division, Lawrence Berkeley National Laboratory, Berkeley, California 94720, USA}
\affiliation{Department of Physics, University of California, Berkeley, California 94720, USA}

\author{Alun Biffin}
\affiliation{Clarendon Laboratory, University of Oxford Physics Department, Parks Road, Oxford OX1 3PU, UK}

\author{Sungkyun Choi}
\affiliation{Clarendon Laboratory, University of Oxford Physics Department, Parks Road, Oxford OX1 3PU, UK}

\author{Roger D. Johnson}
\affiliation{Clarendon Laboratory, University of Oxford Physics
Department, Parks Road, Oxford OX1 3PU, UK}

\author{Radu Coldea}
\affiliation{Clarendon Laboratory, University of Oxford Physics
Department, Parks Road, Oxford OX1 3PU, UK}
\author{Pilanda Watkins-Curry}
\affiliation{Department of Chemistry, The University of Texas at Dallas, Richardson, Texas 75080, USA}

\author{Gregory T. McCandless}
\affiliation{Department of Chemistry, The University of Texas at Dallas, Richardson, Texas 75080, USA}
\author{Julia Y. Chan}
\affiliation{Department of Chemistry, The University of Texas at Dallas, Richardson, Texas 75080, USA}
\author{Felipe Gandara}
\affiliation{Materials Science Division, Lawrence Berkeley National Laboratory, Berkeley, California 94720, USA}
\author{Z. Islam}
\affiliation{Advanced Photon Source, Argonne National Laboratory, Argonne, Illinois 60439, USA}

\author{Ashvin Vishwanath}
\affiliation{Materials Science Division, Lawrence Berkeley National Laboratory, Berkeley, California 94720, USA}
\affiliation{Department of Physics, University of California, Berkeley, California 94720, USA}
\author{Arkady Shekhter}
\affiliation{Los Alamos National Laboratory, Los Alamos, NM 87545, USA}
\author{Ross D. McDonald}
\affiliation{Los Alamos National Laboratory, Los Alamos, NM 87545, USA}
\author{James G. Analytis}
\affiliation{Materials Science Division, Lawrence Berkeley National Laboratory, Berkeley, California 94720, USA}
\affiliation{Department of Physics, University of California, Berkeley, California 94720, USA}

\begin{abstract}

{\bf Spin and orbital quantum numbers play a key role in the physics of Mott insulators, but in most
systems they are connected only indirectly --- via the Pauli
exclusion principle and the Coulomb interaction. Iridium-based
oxides (iridates)
introduce strong spin-orbit coupling directly, such that the Mott
physics has a strong orbital character. In the layered honeycomb
iridates this is thought to generate highly spin-anisotropic
magnetic interactions, coupling the spin orientation to a given spatial
direction of exchange and leading to strongly frustrated
magnetism. Here we report a new iridate
structure that has the same local connectivity as the layered
honeycomb and exhibits striking evidence for highly
spin-anisotropic exchange.  The basic structural units of this material suggest that a new family of three-dimensional structures could exist, the `harmonic honeycomb' iridates, of which the present compound is the first example.  }
\end{abstract}


\maketitle
\section{Introduction}

Quantum spin systems are characterized by small moments where the
spin orientation is decoupled from the crystal lattice, in
contrast to Ising-like spin systems that often apply to higher
spin states. In the Heisenberg model describing spin-isotropic
exchange between neigboring spins, spatial anisotropies of the
exchange suppress long-range order \cite{mermin_absence_1966}, but
do not lead to anisotropy of the magnetic susceptibility. Striking
examples of this are quasi-1D and -2D systems where the exchange
differs by orders of magnitude for neighbors along distinct
crystallographic directions \cite{skomski_simple_2008,
goddard_experimentally_2008}. The spin-orbit interaction
introduces magnetic anisotropy by coupling the spin to the
symmetry of the local orbital environment. Although in
spin-$\sfrac12$ systems the crystal field does not introduce
single-ion anisotropy (due to Kramer's protection of the
spin-$\sfrac12$ doublet), it can --- via spin-orbit --- introduce
spin-anisotropies in the $g$-factor and in the exchange
interactions. The strength of the spin-orbit coupling varies by orders
of magnitude between the $3d$ and $5d$ transition metals. In the
former, quenching of the orbital moment decouples the orbital
wavefunction from the spin, giving a $g$-factor anisotropy that is
typically small and an even smaller spin-anisotropy. For example,
spin-$\sfrac12$ copper in a tetragonal crystal field has a
$g$-factor anisotropy of order $10\%$, whereas the spin-anisotropy
of exchange is of the order of $1\%$
\cite{goddard_experimentally_2008}.

The stronger spin-orbit coupling of the $5d$ transition metals is
known to give rise to larger magnetic anisotropies. In materials
with edge-shared IrO$_6$ octahedra, spin-anisotropy of the
exchange between neighboring effective spin-$\sfrac{1}{2}$ states
is enhanced by the interference of the two exchange paths across
the planar \ioi\, bond. Jackeli and Khaliullin (JK) suggested that
in the honeycomb iridates this may lead to extreme spin-anisotropy
of the exchange coupling, where in the limiting case, the only
non-vanishing interaction is for the spin component normal to the
\ioi\, plane
\cite{Khaliullin2009,Khaliullin2010,chaloupka_zigzag_2013}. In the
honeycomb lattice, such an interaction couples different
orthogonal spin components for the three nearest neighbors; no
single exchange direction can be simultaneously satisfied, leading
to strong frustration. It is the possibility of engineering
spin-anisotropy coupled to spatial exchange pathways that has
spurred intense scientific research, particularly in connection to
the search for quantum spin-liquids
\cite{kitaev_anyons_2006,Khaliullin2009,Khaliullin2010,chaloupka_zigzag_2013}.
However, whether the spin-anisotropic exchange interaction that is
coupled to the \ioi\, bonding plane is realized in such materials
remains an intense subject of scientific debate
\cite{Gegenwart2012, chaloupka_zigzag_2013,choi_spin_2012,
gretarsson_magnetic_2013}, highlighting the need for the discovery
of new materials with related structures and strongly anisotropic
exchange interactions.

\begin{figure}
\includegraphics[width=7cm]{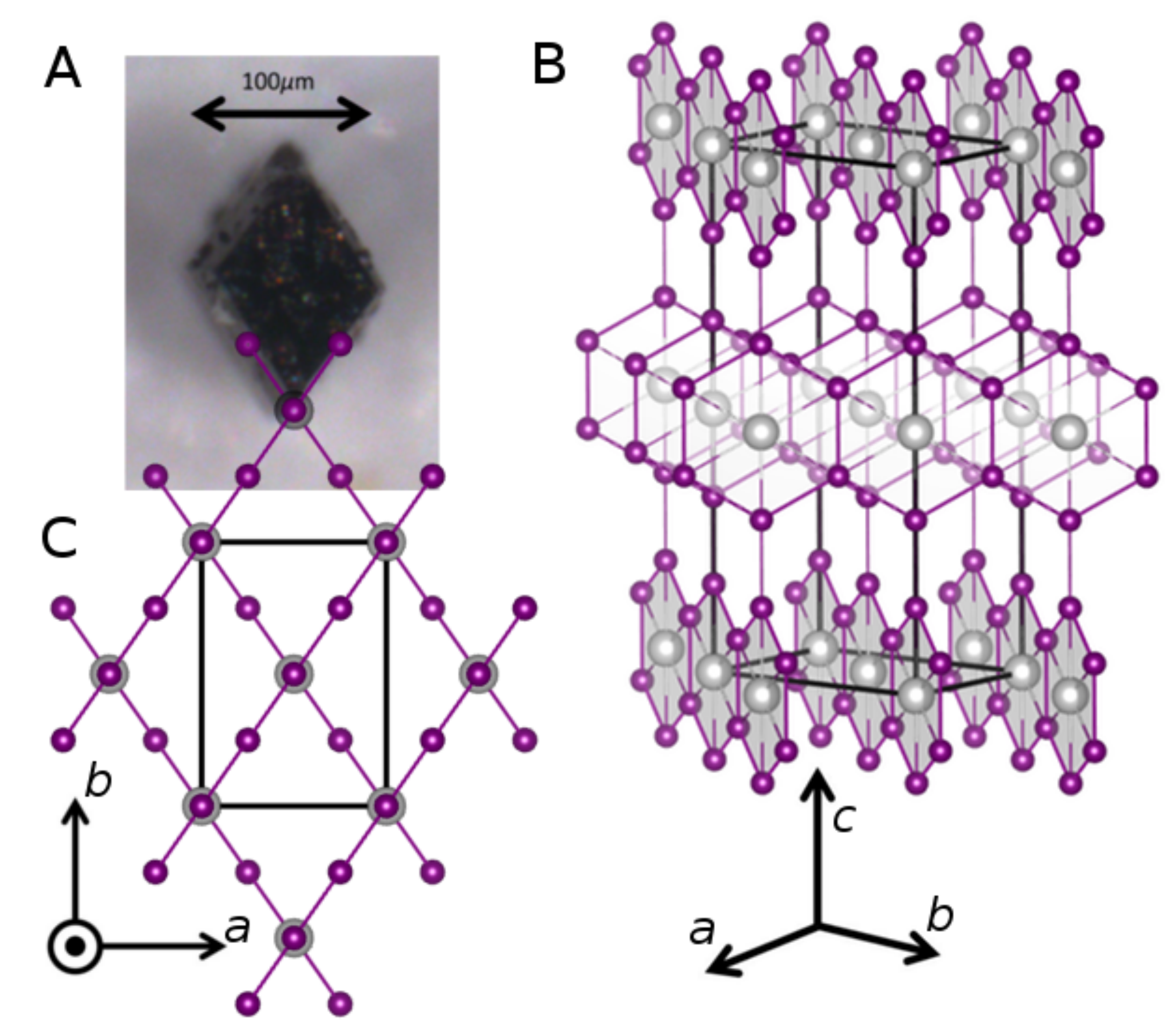}
\caption{{\bf Single crystal of $^\mathcal{H}\langle 1\rangle$-\li213\,
and the Ir lattice structure.} (A) Single crystal oriented to be
parallel to the crystallographic axes shown in (C), (B) 3D view
and (C) projection in the $ab$ plane. In (B) gray shading
emphasizes the Ir (purple balls) honeycomb rows that run parallel
to the $\bm{a}\pm\bm{b}$ diagonals, alternating upon moving
along the $c$-axis. For simplicity only Li ions (grey balls)
located in the center of Ir honeycombs are shown. In (B) and (C)
the rectangular box indicates the unit cell. Comparing (A) and (C)
we note that the $\sim$70$^\circ$ angle between honeycomb rows is
evident in the crystalline morphology.} \label{fig:xtalstruc}
\end{figure}

We have synthesized single crystals of a new polytype of \li213 in
which we reveal the effect of the spin-anisotropy of exchange from
the temperature dependence of the anisotropic magnetic
susceptibility.

\section{Results}
\subsection{Crystal structure}
Single crystals of \li213 were synthesized as described in
Methods. As shown in Figure \ref{fig:xtalstruc}A, the crystals are
clearly faceted and typically around $100 \times 100 \times 200
\mu {\rm m}^3$ in size. In contrast to the monoclinic structure of
the layered iridate, we find that these materials are orthorhombic
and belong to the non-symmorphic space group $Cccm$, with lattice
parameters $a=5.9119(3)$~\AA , $b=8.4461(5)$~\AA ,
$c=17.8363(10)$~\AA ~(see Supplementary Discussion for
details of the crystallography).  The structure (shown in Figure
\ref{fig:xtalstruc}B and C) contains two interlaced honeycomb
planes, the orientation of which alternate along the $c$ axis. The x-ray refinement (see
Supplementary Discussion, Supplementary Figures 1-4 and Supplementary Tables) is consistent with fully
stoichiometric \li213. In this case the Ir oxidation state is
Ir$^{4+}$ ($5d^5$), fixing the effective Ir local moment
$J_{\rm{\rm eff}}=\sfrac12$, which is consistent with the magnetic
properties of our crystals (see Figure \ref{fig:chi}). In
addition, highly-sensitive single-crystal susceptibility and
torque measurements (see below) observe sharp anomalies at the
transition to magnetic order, with no measurable variability in
this transition temperature between many crystals measured,
indicating that the observed magnetic order is well-formed and
intrinsic to the crystals. This suggests that if present, Li
vacancy disorder is small, because such vacancies will to lead
non-magnetic Ir$^{5+}$
$5d^4$\cite{bremholm_nairo3pentavalent_2011}, whose presence is
expected to give rise to  spin-glass behavior
\cite{andrade_magnetism_2013} which we do not observe. Taken
together, our experiments indicate that our crystals are
representative of the high-purity, stoichiometric limit. We denote
the crystal structure \1hh, where $^\mathcal{H}\langle 1\rangle$
refers to the single, complete $\mathcal{H}$oneycomb row.

\begin{figure}[h!]
\includegraphics[width=10cm]{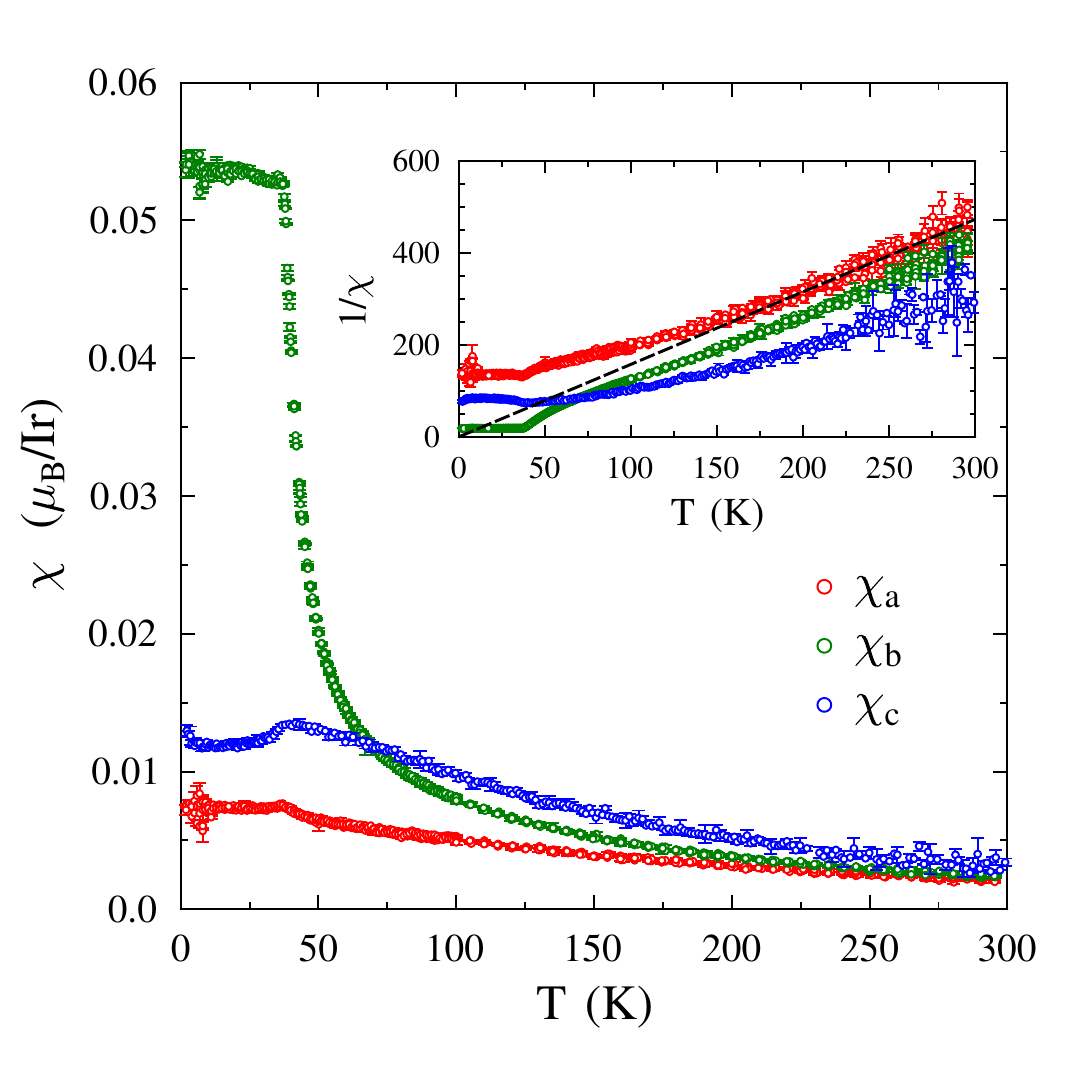}
\caption{{\bf The temperature dependence of the single-crystal
magnetic susceptibility along the three principal crystalline
directions.} The inset shows $1/\chi$ for all three axes $\chi_a$,
$\chi_b$, and $\chi_c$. The dashed line indicates the slope of the
inverse Curie-Weiss  susceptibility for a paramagnet with
effective moment of $\mu_{\rm eff}=1.6\mu_B$, close to that
expected of an Ir $J_{\rm eff}=\sfrac12$ state if $g$-factor
anisotropy is ignored. All three components of susceptibility show
strong deviation from Curie-Weiss behavior as a function of
temperature.}
\label{fig:chi}
\end{figure}

\subsection{High temperature magnetic anisotropy}
As can be seen in Figure \ref{fig:chi}, the raw magnetic
susceptibility shows a magnetic anomaly at 38~K, most likely
reflecting the bipartite nature of the structure, which alleviates
the magnetic frustration. Due to the smallness of our samples and
sensitivity to sample misalignment, the anisotropy at high
temperatures could not be quantitatively resolved to high accuracy
using SQUID magnetometry. To do so, we utilized torque
magnetometry, which exclusively probes magnetic anisotropy (see
discussion in Supplementary Discussion) and is sufficiently sensitive to measure
single crystals of $\sim 10\mu$m dimensions. Torque magnetometry
was measured by attaching an oriented single crystal to a
piezoresistive micro-cantilever \cite{ohmichi_torque_2002} that
measures mechanical stress as the crystal flexes the lever to try
to align its magnetic axes with the applied field.  The mechanical
strain is measured as a voltage change across a balanced
Wheatstone Bridge and can detect a torque signal on the order of
$10^{-13}$~Nm. The lever only responds to a torque perpendicular
to its long axis and planar surface. As a result, the orientation
of the crystal on the lever (determined by x-ray measurements and
the diamond shaped morphology) defines the plane of rotation in
field and which principal components of anisotropy, $\alpha_{ij}$
($i,j\in a,b,c$) are measured. To achieve this the cantilever was
mounted on a cryogenic goniometer to allow rotation of the sample
with respect to magnetic field without thermal cycling. The low
temperature anisotropy was confirmed on several similar sized
single crystals. To measure $\alpha_{ij} = \chi_i-\chi_j$ between
1.5~K and 250~K, three discrete planes of rotation for the same
crystal were used.

A magnetically anisotropic material experiences a torque
when its magnetization is not aligned with the applied magnetic
field; the deflection of the cantilever in a uniform magnetic
field is hence a direct measure of the magnetic anisotropy,
$\vec{\tau}=\vec{M}\times\vec{H}$.
At small fields, where the magnetic response is linear, the
magnetic anisotropy is captured by a susceptibility tensor
$M_i=\chi_{ij}H_j$. For an orthorhombic crystal, the magnetic axes
coincide with the crystallographic directions, defining
$\chi_{a,b,c}$. For example, for rotations in the
$\bm{{b}}$-$\bm{{c}}$ plane, the anisotropic magnetization
$(M_b,M_c)=(\chi_b H_b, \chi_c H_c)$ creates a torque
\begin{align}
\tau_a =\frac{(\chi_b-\chi_c)H^2\text{sin}2\theta}{2}
\label{eq:sin2theta}
\end{align}
where $\theta$ is the angle between a crystallographic axis ($c$
in this case) and the applied magnetic field.

\begin{figure}[h!]
\includegraphics[width=14cm]{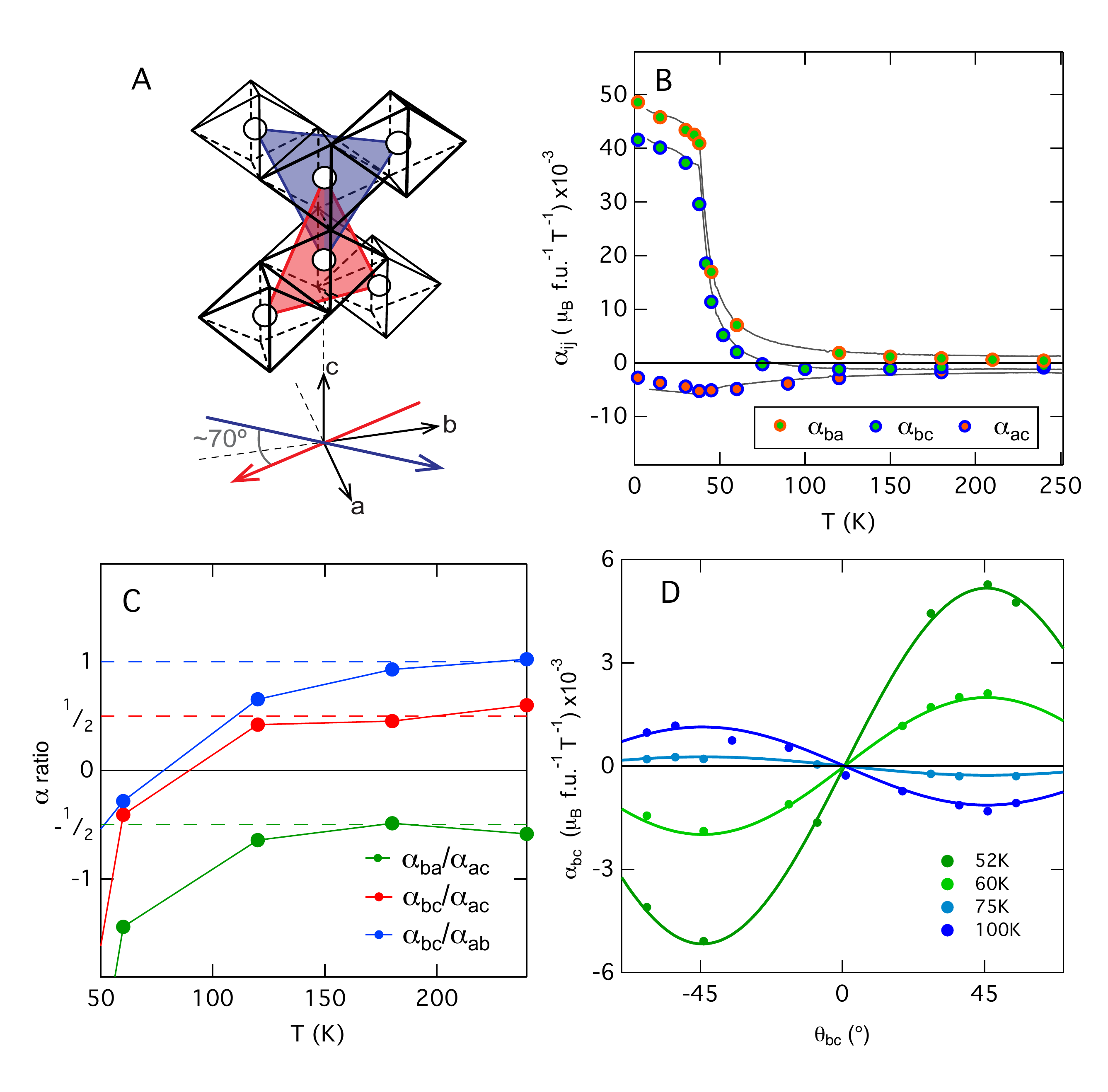}
\caption{{\bf Temperature dependence of the magnetic anisotropy.}
(A) Each Ir is surrounded by one of two planar, triangular
environments indicated by blue and red shaded triangles, located
at $\sim 35^\circ$ either side of the $b$-axis. (B) The anisotropy
of the magnetic susceptibility as measured by torque and the
differences in (SQUID) susceptibilities (grey lines) are shown as
a function of temperature for all three crystallographic
directions.  An anomaly indicates the onset of magnetic order at
$T_{\rm N}=38$~K. (C) The ratios of the anisotropic susceptibility
tend to simple fractional values dictated by the $g$-factor
anisotropy of the local planar iridium environment. (D)
sin$(2\theta)$ fits to the anisotropy $\alpha_{bc}$ illustrating
the change of sign at $\sim 75$K.  } \label{fig:MhighT}
\end{figure}

The geometry of the lattice is intimately connected to the magnetic anisotropy. Specifically, we note that the
angle $\phi_0$ between the honeycomb strips (see Figure \ref{fig:xtalstruc}C) is fixed by the
geometry of the edge shared bonding of the IrO$_6$ octahedra (see
Figures \ref{fig:MhighT}A). For cubic
octahedra $\cos\phi_0=\sfrac13$, namely $\phi_0\approx 70^\circ$,
as shown in Figure \ref{fig:MhighT}A. The magnetic axes can be uniquely identified from a complete
angular dependence of the torque measurements, with the amplitude
of the $\text{sin}2\theta$ dependence being proportional to the
magnetic anisotropy $\alpha _{ij}$. The observed magnetic axes are
independent of temperature between 300~K and 1.5~K. The magnetic
anisotropy, shown as data points in Figure \ref{fig:MhighT}B
agrees well with the differences in the low temperature
susceptibility data (grey lines in Figure \ref{fig:MhighT}B).  At
temperatures that are high relative to the exchange interaction
energy scale, we expect that only the $g$-factor affects the
magnetic anisotropy.  We find that the ratio of the anisotropic
susceptibilities $\alpha_{ij}/\alpha_{jk}$ asymptotically approach
simple fractions at high temperature (above $\sim$100~K, see
Figure \ref{fig:MhighT}C). Specifically, each Ir is in a
three-fold local planar environment with (almost) equidistant
neighbors and thus the Ir $g$-factor anisotropy can be captured by
ascribing each honeycomb plane susceptibility components parallel,
$\chi_\parallel$ and perpendicular, $\chi_\perp$ to the plane
(consider Figure \ref{fig:MhighT}A). This uniaxial local iridium
environment combined with the relative orientation of the iridium
planes, $\cos\phi_0=\sfrac13$, constrains the three components of
susceptibility at high temperature to be equally spaced;
$2\chi_b=\chi_a+\chi_c$ (see Supplementary Discussion)
and the anisotropy ratios to be
$\alpha_{ba}/\alpha_{ac}=-\sfrac12,\alpha_{bc}/\alpha_{ac}=\sfrac12,\alpha_{bc}/\alpha_{ab}=1$,
just as we observe. This observation places constraints on the
ordering of the principal components of the $g$-factor at all
temperatures.

\subsection{Reordering of the principal magnetic axes}
The striking reordering of the principal components of
susceptibility revealed in torque and SQUID magnetometry, is
associated with a strong deviation from Curie-Weiss behavior as
the temperature is lowered: $\alpha_{bc}$ changes sign at
$T\approx 75$~K (Figure \ref{fig:MhighT}~D and Supplementary Figures 5-6). This is in stark
contrast to spin-isotropic Heisenberg exchange systems where the
low temperature susceptibility reflects the $g$-factor anisotropy
observed at high temperatures, even in the presence of
spatially-anisotropic exchange \cite{goddard_experimentally_2008}.
The change of sign of $\alpha_{bc}$ arises because $\chi_b$
softens, becoming an order of magnitude greater than $\chi_a$ and
$\sim5\times\chi_c$ (Figure \ref{fig:chi} and \ref{fig:MhighT}B).
As a result, the susceptibility cannot be parameterized by a
Curie-Weiss temperature: the linear extrapolation of all three
components of inverse susceptibility to the temperature axis
depends strongly upon the temperature range considered. Between
50--150~K the extrapolation of all three components of inverse
susceptibility is negative, consistent with the absence of net
moment in the ordered state. However, at higher temperatures
(200--300~K) the inverse susceptibilities $1/\chi_b,1/\chi_c$
extrapolate to positive temperature intercepts (see Figure
\ref{fig:chi}) indicating a ferromagnetic component to the
interactions. Above 200~K, $1/\chi$ the Curie-Weiss slope gives
$\mu_{\rm eff}\approx1.6\mu_{\rm B}$, consistent with a $J_{\rm
eff}=\sfrac12$ magnetism.

The observed ten fold increase in $\chi_b$ cannot be driven by the
$g$-factor of the local iridium environment, whose geometric
constraints are temperature independent (see Supplementary
Discussion). The temperature dependence of
$\chi_b$ must therefore arise from spin-anisotropic exchange. We
note that all the $c$-axis bonds have the \ioi\, plane normal to
the $b$-axis, whether they preserve or rotate between the two
honeycomb orientations (see the full structure in Supplementary Figure 3A and a schematic in Figure \
\ref{fig:MlowT}A - green shading indicate the \ioi\, planes). This
is the only \ioi\, plane that is normal to a crystallographic
axis. This coupling of the spin-anisotropy to the structure,
provides evidence for spin-anisotropic exchange across the
$c$-axis links, and by extension should be present in all \ioi\
exchange paths. This likely arises from the interfering exchange
mechanism suggested by JK in the context of the Kitaev model (see
Supplementary Discussion).

\begin{figure}
\includegraphics[width=16cm]{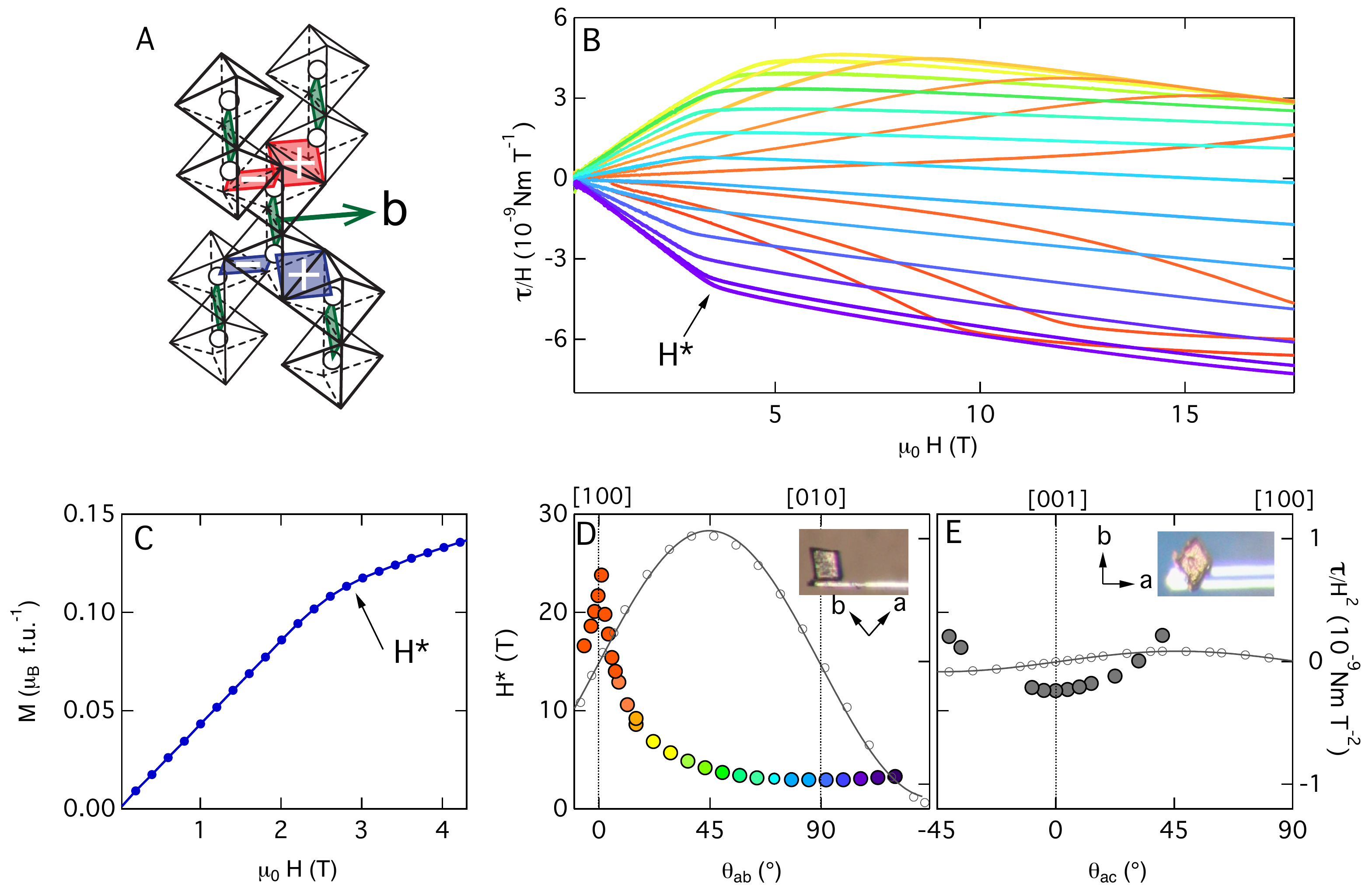}
\caption{{\bf Low temperature magnetic properties of the \1hh.}
(A) The \ioi\, planes defining three orthogonal directions of the
spin-exchange, one parallel to $\hat{\bm{b}}$ and the other two
parallel to $\hat{\bm{a}}\pm\hat{\bm{c}}$, labelled $+$ and $-$
($\hat{\bm{a}}$ is the unit vector along $\bm{a}$). This connects
to the notation used to describe the Kitaev Hamiltonian in SI III.
(B) Torque signal $\tau$ divided by the applied magnetic field $H$
at a temperature of 1.5~K, illustrating a linear low-field
dependence and a kink at $H^*$, which is strongly angle dependent
(colors correspond to angles shown in (D)). (C) Magnetization vs
magnetic field applied along the $b$-axis at a temperature of
15~K. (D) $\&$ (E) The angle dependence $\theta_{ab/ac}$ of the
kink field $H^*$ of the ordered state (full circles, left axes)
with respect to the crystallographic axes $a,b$ and $c$.  $H^*$ is
correlated to the magnetization anisotropy $\alpha_{ij}$ (open
circles, right axes) indicating a common moment at $H^*$ in all
field orientations.} \label{fig:MlowT}
\end{figure}

\subsection{Low temperature magnetic properties}
The softening of $\chi_b$ is truncated at 38~K by a magnetic
instability. Within the ordered state, the magnetization increases
linearly with applied field (Figure \ref{fig:MlowT}C, $\tau/H$
in \ref{fig:MlowT}B and Supplementary Figure 6).  At sufficiently high magnetic fields $H^*$,
the magnetization kinks abruptly.  This corresponds to an induced
moment of $\approx 0.1\mu_B$.  Above $H^*$, the finite torque
signal reveals that the induced moment is not co-linear with the
applied field, consistent with the finite slope observed at these
fields in Figure \ref{fig:MlowT}C. This shows that in the phase
above $H^*$ the induced magnetization along the field direction is
not yet saturated (the value is well below the expected saturated
Ir moment of $\sim$1$\mu_B$ for $J_{\rm eff}=\sfrac12$). The
angular dependence of both the slope of the linear regime and the
kink field $H^*$, exhibit an order of magnitude anisotropy with
field orientation (Figure \ref{fig:MlowT}D and \ref{fig:MlowT}E).
Such strong anisotropy in a spin-$\sfrac12$ system highlights the
strong orbital character arising from the spin-orbit coupling,
again in contrast to spin-$\sfrac12$ Heisenberg
anti-ferromagnetism \cite{goddard_experimentally_2008}.

\section{Discussion}
\begin{figure}
\includegraphics[trim=0 200 0 0,clip,width=\textwidth]{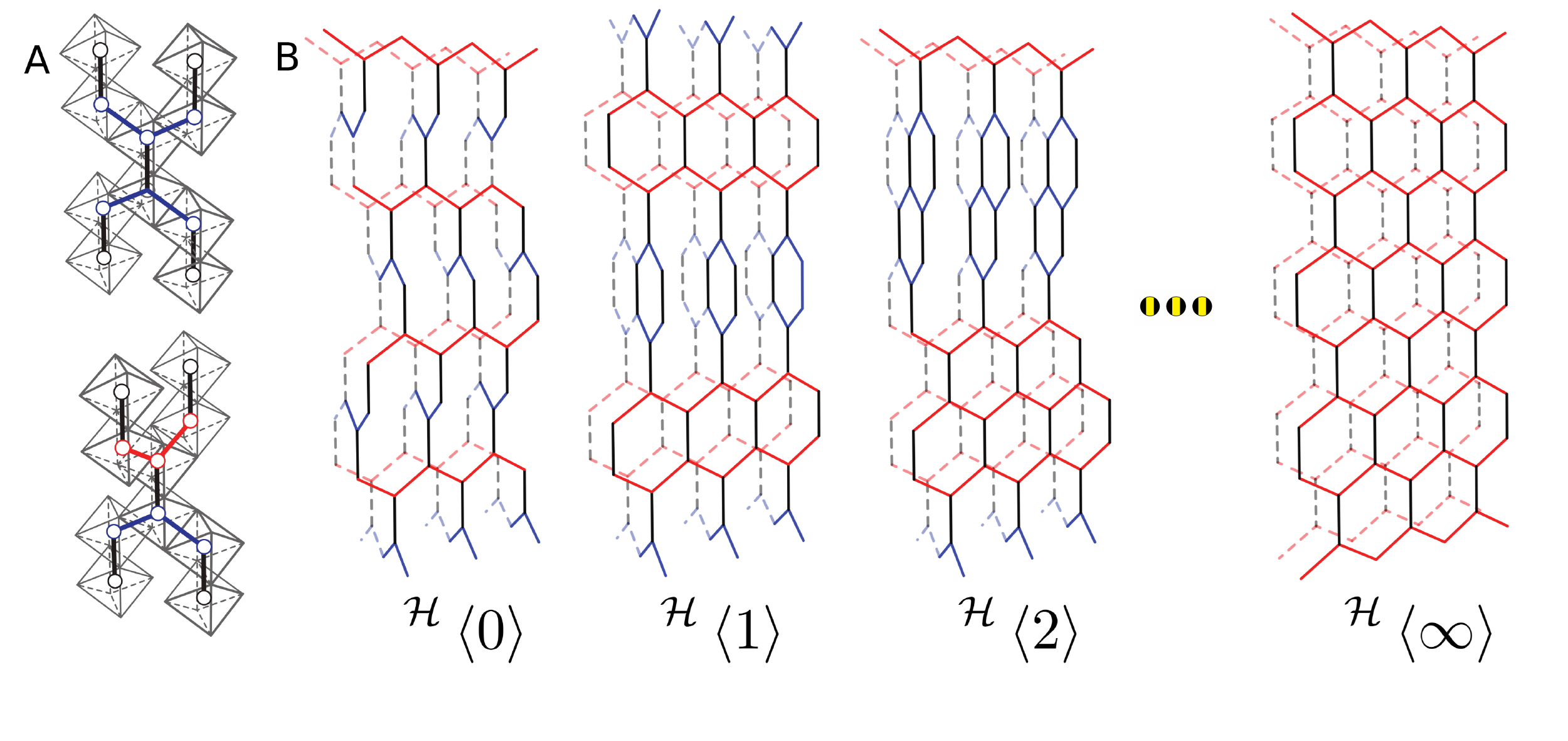}
\caption{{\bf Introducing the harmonic honeycomb series.} (A) Two kinds
of $c$-axis bonds (black links) in the harmonic honeycomb family
\nhh\, are shown, one linking within a honeycomb plane (for
example blue to blue, top) and one that rotates between honeycomb
planes (for example red to blue, bottom). For undistorted
octahedra, these links are locally indistinguishable, as can be
observed by the local coordination of any Ir atom (also see Figure
\ref{fig:MhighT}A). (B) These building blocks can be used to
construct a series of structures. The end members include the
theoretical $N=0$ `hyper-honeycomb'
\cite{mandal_exactly_2009,kimchi_three_2013,lee_heisenberg-kitaev_2013}
and the $N=\infty$ layered honeycomb
\cite{omalley_structure_2008}. Here $N$ counts the number of
complete honeycomb rows in a section along the $c$-axis before the
orientation of the honeycomb plane switches.} \label{fig:harmonic}
\end{figure}

There is a very interesting connection between the layered
honeycomb \li213\, and the polytype studied here. The \1hh\, is
distinguished by its $c$-axis bond, which either preserves or
rotates away from a given honeycomb plane (see Figure
\ref{fig:harmonic}A and Supplementary Figure 7); in the case that all the bonds preserve the
same plane, the resulting structure is the layered honeycomb
system. Further polytypes can be envisioned by tuning the $c$-axis
extent of the honeycomb plane before switching to the other
orientation (see Figure \ref{fig:harmonic}B). We denote each
polytype $^\mathcal{H}\langle N\rangle$-\li213, where
$^\mathcal{H}\langle N\rangle$ refers to the number of complete
honeycomb rows (see Figure \ref{fig:harmonic}B and Supplementary Figure 8), and the family as
the ``harmonic"-honeycombs, so named to invoke the periodic
connection between members. The layered compound,
$^\mathcal{H}\langle\infty\rangle$-\li213
\cite{omalley_structure_2008} and the hypothetical hyper-honeycomb
structure,  $^\mathcal{H}\langle
0\rangle$-\li213\,\cite{kimchi_three_2013} are the end members of
this family (see also SI IV). The edge-sharing geometry of the
octahedra preserves the essential ingredients of the Kitaev model
and this is universal for this family of polytypes. Each structure
is a material candidate for the realization of a 3D spin liquid in
the pure Kitaev limit (see Supplementary Discussion
and for \0hh\, see Refs.
\cite{mandal_exactly_2009,lee_heisenberg-kitaev_2013,kimchi_three_2013}).

Finally, we speculate on the consequences and feasibility of
making other members of the \nhh\, family. Both the layered
$^{\mathcal{H}}\langle\infty\rangle$-\li213 and the
$^{\mathcal{H}}\langle1\rangle$-\li213 are stable structures,
implying that intermediate members may be possible under
appropriate synthesis conditions. The building blocks shown in
Figure \ref{fig:harmonic}A connect each member of the harmonic
honeycomb series in a manner that is analogous to how corner
sharing octahedra connect the Ruddlesden-Popper (RP) series.
Indeed, despite the fact that members of the RP family are locally
identical in structure, they exhibit a rich variety of exotic
electronic states; including superconductivity and ferromagnetism
in the ruthenates \cite{mazin_competitions_1999,
grigera_magnetic_2001}, multiferroic behavior in the titanates
\cite{benedek_hybrid_2011}, collosal magnetoresistance in the
manganites \cite{mihut_physical_1998} and high temperature
superconductivity in the cuprates \cite{leggett_cuprate_1999}. The
harmonic honeycomb family is a honeycomb analogue of the RP
series, and its successful synthesis could similarly create a new
frontier in the exploration of strongly spin-orbit coupled Mott
insulators.

\section{Methods}
\label{sec:synthesis}
\subsection{Synthesis}
Powders of IrO$_2$ (99.99$\%$ purity, Alfa-Aesar) and Li$_2$CO$_3$
($99.9\%$ purity, Alfa-Aesar) in the ratio of 1:1.05, were reacted
at 1000$^\circ$C, then reground and pelletized, taken to
1100$^\circ$C and cooled slowly down to 800$^\circ$C. The
resulting pellet was then melted in LiOH in the ratio of 1:100
between 700-800$^\circ C$ and cooled at 5$^\circ$C/hr to yield
single crystals of \1hh. The crystals were then mechanically
extracted from the growth. Single crystal x-ray refinements were
performed using a Mo-source Oxford Diffraction Supernova
diffractometer.Please see Supplementary Discussion for
a detailed analysis.

\subsection{Magnetic measurements}
Two complementary techniques were used to measure the magnetic
response of single crystals of $^\mathcal{H}\langle 1
\rangle$-\li213; a SQUID magnetometer was employed to measure
magnetization and a piezoresistive cantilever to directly measure
the magnetic anisotropy. The magnetization measurements were
performed in a Cryogenic S700X. Due to the size of the single
crystals, the high temperature magnetization was near the noise floor of the
experiment. Nevertheless, SQUID measured anisotropies at high
temperatures were close to those measured by torque, with absolute
values of $\chi_a (300$K$)=0.0021 \mu_{\rm B}/$T f.u. and
$\chi_b=0.0024 \mu_{\rm B}$/T f.u.. Curie-Weiss fits to the linear
portion of the susceptibility yielded an effective moment of
$\mu_{\rm eff}=1.6(1) \mu_{\rm B}$, consistent with $J_{\rm
eff}=\sfrac12$ magnetism. However, the SQUID resolution was not
adequate to determine the susceptibilities anisotropy at high
temperature to the accuracy we required (see Figure \ref{fig:chi}
). To resolve the magnetic anisotropy throughout the entire
temperature range, we employed torque magnetometry, where a single
crystal  could be precisely oriented. Although the piezoresistive
cantilever technique is sensitive enough to resolve the anisotropy
of a$\sim50 \mu\text{m}^3$ single crystal, and hence ordering of
susceptibilities at high temperature, the absolute calibration of
the piezoresistive response of the lever leads to a larger
systematic error than in the absolute value of the susceptibility
measured using the SQUID at low temperature. To reconcile these
systematic deficiencies in both techniques, the torque data was
scaled by a single common factor of the order of unity, for all
field orientations and temperatures, so as to give the best
agreement with the differences between the low temperature
susceptibilities as measured using the SQUID. The rescaled torque
data was thus used to resolve the magnetic anisotropy at high
temperature where the susceptibility is smallest.

Torque magnetometry was measured on a $ 50 \times 100 \times 40
\mu {\rm m}^3$ single crystal ($5.95\times 10^{-9}$ mol Ir)
employing a piezoresistive micro-cantilever
\cite{ohmichi_torque_2002} that measures mechanical stress as the
crystal flexes the lever to try to align its magnetic axes with
the applied field.  The mechanical strain is measured as a voltage
change across a balanced Wheatstone Bridge and can detect a torque
signal on the order of $10^{-13}$~ Nm.  Torque magnetometry is an
extremely sensitive technique and is well suited for measuring
very small single crystals.  The cantilever was mounted on a
cryogenic goniometer to allow rotation of the sample with respect
to magnetic field without thermal cycling. The lever only responds
to a torque perpendicular to it's long axis and planar surface,
such that the orientation of the crystal on the lever and the
plane of rotation in field could be chosen to measure the
principal components of anisotropy, $\alpha_{ij}$. The low
temperature anisotropy was confirmed on several similar sized
single crystals. However to measure $\alpha_{ij} = \chi_i-\chi_j$
between 1.5~K and 250~K, three discrete planes of rotation for the
same crystal were used. When remounting the sample to change the
plane of rotation, care was taken to maintain the same center of
mass position of the crystal on the lever to minimize systematic
changes in sensitivity. Magnetic fields were applied using a 20~T
superconducting solenoid and a 35~T resistive solenoid at the
National High Magnetic Field Laboratory, Tallahassee, FL.

\section{Acknowledgements}

Synthesis and discovery of this compound was supported by the U.S. Department of Energy, Office
of Basic Energy Sciences, Materials Sciences and Engineering
Division, under Contract No. DE-AC02-05CH11231. Magnetization measurements performed by TES and NPB were support by the Laboratory Directed Research and
Development Program of Lawrence Berkeley National Laboratory under
U.S. Department of Energy Contract No. DE-AC02-05CH11231. TES also acknowledges
support from the National Science Foundation Graduate Research Fellowship
under Grant No. DGE 1106400. JYC acknowledges NSF-DMR 1063735 for
support. RMcD acknowledges support from BES-`Science of 100
tesla'. Work at Oxford was supported by the EPSRC (UK) grant
EP/H014934/1. The work at the National High Magnetic Field
Laboratory is supported via NSF/DMR 1157490.

\section{Author contributions}
JGA synthesized the crystals. JGA, RDM conceived of the
experiment. KAM, RDM, AS and JGA performed the torque magnetometry
experiments. NPB, JGA and TES performed the SQUID magnetometry
experiments. AV, IK and AS contributed to the theory. AB, SC, RDJ
and RC solved the crystal structure from single-crystal x-ray
diffraction measurements and wrote the crystallography section of
the paper. PWC, GTM, FG and JYC performed additional x-ray
diffraction measurements. ZI measured changes in the low
temperature crystallographic parameters below $T_{\rm N}$. All authors contributed to the writing of the manuscript.

\newpage

%

\end{document}